\documentclass[reprint,amsmath,amssymb,aps,nofootinbib]{revtex4-1}
\usepackage{float}
\usepackage{bm}
\usepackage{graphicx}
\usepackage[colorlinks, linkcolor= blue, citecolor = blue, urlcolor=blue]{hyperref}
\usepackage{dcolumn}

\usepackage{soul}

\expandafter\ifx\csname package@font\endcsname\relax\else
 \expandafter\expandafter
 \expandafter\usepackage
 \expandafter\expandafter
 \expandafter{\csname package@font\endcsname}%
\fi
\usepackage{times}
\begin{document}

\preprint{APS/123-QED}

\title{Selective Spin Wave Non-reciprocity in Engineered Chiral Magnonic Crystal without Dzyaloshinskii-Moriya Interaction}
\author{Diksha Prajapati$^{\rm{1}}$ and Chandrima Banerjee$^{\rm{1}}$}

\email{cbanerjee@iitk.ac.in}
\affiliation{$^{\rm{1}}$ Department of Physics, Indian Institute of Technology, Kanpur, Uttar Pradesh-208016, India
}

%\date{\today}

\begin{abstract}
Chirality is pivotal in magnonics, particularly for achieving spin wave non-reciprocity which is critical in advancing spin wave based communication and logic operations. In general, chirality in magnetic systems is realized through the interfacial antisymmetric exchange interaction, namely, the Dzyaloshinskii-Moriya Interaction (DMI), which is an intrinsic phenomenon occurring at the buried interface of ferromagnet and heavy metal. In this work, using micromagnetic simulations, we present a new route to achieve spin wave non-reciprocity using an engineered chiral magnonic crystal, where we artificially created in-plane twisted spin textures by imposing certain handedness in the structural geometry. The manipulation of the relative arrangement of these chiral enantiomers led to spin wave non-reciprocity, which could be tuned by selecting different channels within the crystal, with additional versatility in the resonant mode frequency. Further adjustment of the external magnetic field strength confirms that the spin wave directionality originates from the spin texture. This enables additional control over the band gap and mode propagation characteristics by external field tuning. Our work provides a novel and innovative route to simultaneously access opposite spin wave directions in consecutive channels, which makes this system competent for real on-chip magnonic diode, with predefined paths for counter-propagating information. 
\end{abstract} 
\maketitle
\section{Introduction}
Symmetry plays a central role in maintaining uniformity in the fundamental laws of nature. Simultaneously, symmetry breaking in the presence of interactions and boundary conditions can give rise to a plethora of rich and complex behaviour that enriches the understanding of fundamental processes, phase transitions, and emergent phenomena across various scientific disciplines \cite{r1_broken_symmetries,r2_SOS_symmetry_operational_similarity}. In condensed matter systems, the competing anisotropic interactions arising from broken symmetry can favour non-reciprocal properties in the elementary excitations of the quasiparticles. The non-reciprocity is the characteristics of waves to change their properties, such as frequency, amplitude, group velocity etc., when the direction of the wave propagation is reversed \cite{r1_broken_symmetries}. It is a key ingredient in wave based signal processing and computing to design isolators \cite{r3_nonlinear_optical_isolators}, circulators \cite{r4_elastic_wave_circulator}, rectifiers \cite{r5_rectifier_wave_based}, switches and diode emitters \cite{r6_elastic_diode,r7_controllabe_transmission_switch}, which are immune to backscattering and can prevent the cross-talk between receivers and transmitters \cite{r8_magnetic_free_nonreciprocity,r9_Non-reciprocal_photonics}. In magnetic systems, the collective precession of electron spins which can propagate in magnetic materials without the motion of electrons in the form of magnons or spin waves (SWs) are promising for future magnonic devices \cite{r10_Magnonics_SW_on_nanoscale,r11_building_blocks_of_magnonics,r12_Magnon_spintronics,r13_SW_diode_and_circulator}. The field of magnonics exploits SWs to carry and process information which feature low power consumption as compared to traditional electronic devices with high signal processing speed (GHz to THz) \cite{r11_building_blocks_of_magnonics,r12_Magnon_spintronics}. In addition, the flexibility to achieve desired spin wave properties by controlling the bias magnetic field or the system architecture has boosted the research interest and opened new avenues for high speed data processing devices. In this context, the non-reciprocity of SWs is of particular interest and achieving non-reciprocity in the spin wave propagation in a reliable and tunable manner is of high relevance for magnon based logic operations. \par Apparently, non-reciprocity is inherent in magnonics \cite{r14_damon_eshbach_mode}, since the amplitude of the propagating dipolar surface SW mode is primarily confined to one surface of magnetic film. However, this amplitude asymmetry is not accompanied by non-reciprocity in frequency. The frequency non-reciprocity rather could be achieved when the geometry of the magnetic system lacks inversion symmetry, such as different anisotropies at the bottom and top surfaces \cite{r15_SW_nonreciprocity_metallized_MC,r16_frequency_nonreciprocity_SSW}, magnetization grading \cite{r17_magnetization_grading} and dipolarly coupled structures \cite{r18_dipolarly_coupled_FM_bilayer}. In addition, the magneto dipolar interactions as a source for chirality in domain walls and 3D curved surfaces were also reported to favour SW non-reciprocity \cite{r19_curvature_induced_asymmetric_dispersion,r20_unidirectional_SW_channeling_domain_walls}. Nevertheless, the most extensively studied phenomena to achieve SW non-reciprocity in ultrathin films is the interfacial Dzyaloshinskii-Moriya interaction (iDMI) \cite{r21_dzyaloshinsky,r22_DMI_induced_by_symmetry_breaking,r23_DMI_Pt/Co/Ni_film,r24_i-DMI_asymmetric_SW_W/CoFeB/SiO2_heterostructure,r25_i-DMI_Pt/CoFeB_films}, namely the antisymmetric exchange interaction arising at the interface of a ferromagnetic film and a heavy metal with high spin-orbit coupling. This phenomenon leads to chiral spin textures such as skyrmions and chiral domain walls \cite{r26_chiral_magnetic_domian_wall,r27_magnetic_skyrmions,r28_Skyrmions_on_track,r29_chiral_spin_torque_domain_walls,r30_Room-temperature_chiral_magnetic_skyrmions} at the interface, which can give rise to non-reciprocal SW propagation \cite{r31_nonreciprocal_SW_channeling_DMI,r32_magnonic_waveguides_based_on_domain_walls,r33_probing_DMI_ultrathin_films_BLS,r34_SW_dynamics_skyrmion_strings}. In all these cases the SW non-reciprocity is an internal property, which largely depends on material parameters, the type and quality of interface, internal spin texture etc. Moreover, the iDMI effect is relatively weak, so that it is often either neglected or even unnoticed whatsoever, limiting its practical usage. \par In this article, using micromagnetic simulation framework, we introduce an intriguing new platform of chirality and symmetry breaking for attaining non-reciprocal SW propagation by using artificial chiral magnonic crystal (CMC). Magnonic crystals (MCs) are externally designed metamaterials that feature periodic changes in their magnetic properties, enabling the modulation of SWs \cite{r35_SW_in_microstructured_MC,r36_bicomponent_MC_forbidden_band_gaps,r37_pseudo_1D_MC,r38_YIG_based_MC,r39_Cubic_anisotropy_reconfigurable_MC,r40_AC_tunable_MC}. We considered a MC which consists of a periodic arrangement of   magnetic nanostructures, where the structural geometry of each element has certain handedness. This gives rise to chiral spin textures, which, upon periodically altering the handedness, induces asymmetric spin wave propagation. Simulation of the magnonic band structure and spatial profiles of the detected modes showed that the preferred SW propagation direction can be chosen by addressing different channels within the crystal, which is further tunable by selecting the relative orientation of the left handed and right handed enantiomers, the resonant frequency and applied magnetic field. This possibility to simultaneously choose opposite wave vector directions in alternate nanochannels without flipping the bias magnetic field is practically more feasible in terms of a real on-chip magnonic circuit, where all the processing units are integrated along with a predefined path for data flow. This geometry controlled SW propagation tuning was not demonstrated in the chiral magnonic crystals reported earlier, which was formed by a periodically modulated iDMI \cite{r41_experimental_flat_bands_1D_MC,r42_flatbands_indirect_bandgaps_induced_by_periodic_DMI,r43_omnidirectional_flatbands_CMC}. Our findings show that this system may serve as a possible SW diode or rectifier, which are important components for magnonic devices.

\maketitle
\section{Concept and methodology}
\begin{figure}[t]
\centering
\includegraphics[width=\columnwidth]{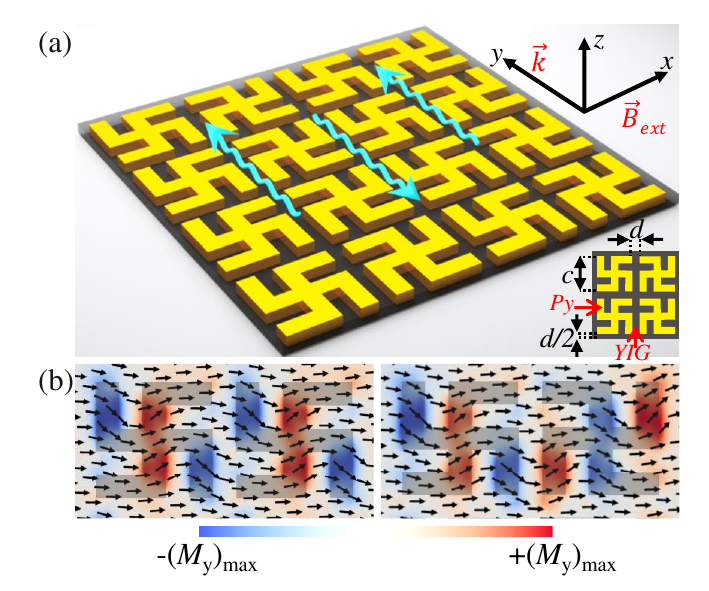}
\caption{(a) Schematic of chiral magnonic crystal with alternate handedness. A typical trend of non-reciprocal spin wave propagation is shown by the arrows. The inset presents the notations for the structural parameters. (b) Static magnetic configurations after applying an external magnetic field of 80 mT to CMC (left panel) and CMCAH (right panel). The spatial variation of $M_{y}$ is shown by the colormap and the elements are indicated by the grey shade as a guide to eye.}
\label{FIG. 1.}
\end{figure}

The chiral magnonic crystal with alternate handedness (CMCAH) under consideration is depicted schematically in Fig.\ref{FIG. 1.}a. The distinctive chiral behaviour arises from the orthogonal arrangement of the arms with respect to each other within one element, where the sense of rotation defines the handedness (left or right). The CMC consists of uniform arrangement of these elements with same handedness. In case of CMCAH, the alternate columns along y-direction have opposite chirality, thus are mirror image of each other. We have considered a bi-component planar crystal where the chiral elements made of permalloy are embedded into YIG matrix, forming a single layer. The thickness of this composite structure is 20nm. \par Micromagnetic simulations were performed using GPU enabled finite-difference method based micromagnetic simulation package MuMax3 \cite{r44_design_verification_MuMax3}. These simulations were employed to explore the magnetization dynamics in both CMC and CMCAH by solving the Landau-Lifshitz-Gilbert (LLG) equation, given as
\[ \frac{\partial \vec{M}(\vec {r},t)} {\partial t} = - \gamma \vec{M}(\vec{r},t) \times \vec{H}_{eff} + \frac{\alpha}{M{s}} (\vec{M}(\vec{r},t) \times \frac{\partial \vec{M}(\vec {r},t)}{\partial t})\]
Where $\vec{\textit{M}}(\vec{\textit{r}},t)$ is the magnetization at a point $\textit{r}$ at time t, $\gamma={\textit{g}\mu_{B}}/{\hbar}$ is gyromagnetic ratio where \textit{g} is Lande g-factor, $\mu_{B}$ is Bohr magnetron and $\hbar$ is reduced Plank’s constant. Gilbert damping constant is given by $\alpha$ and ${H_{eff}}$ is effective magnetic field which includes contributions from exchange, dipolar, zeeman interactions as well as magnetic anisotropy and could be expressed as
\[\vec{H}_{eff}= \vec{H}_{exchange}+\vec{H}_{dipolar}+\vec{H}_{zeeman}+\vec{H}_{anisotropy}\]
The overall dimension of the sample was considered to be 1.2\textmu m$\times9.6$\textmu m$\times$20nm in the x, y and z-directions respectively where the sample was discretized into a total of $384\times3072\times1$ cells, resulting in a cell size of $3.125\times3.125\times20\text{nm}^{3}$. The dimension of each element \textit{c} (See inset of Fig. \ref{FIG. 1.}a) is 250 nm whereas the separation between two consecutive elements \textit{d}, which  is also the width of each arm, is 50 nm (lattice constant \textit{a} = 300nm). The considered material parameters for Py (YIG) were: saturation magnetization $M_{s}=800$ kA/m (143.2 kA/m) and exchange constant $A=13\times10^{-12}\text{J/m}$ $(3.3\times10^{-12}\text{J/m})$. Gilbert damping coefficient was set to be 0.005. Periodic boundary conditions were employed along x and y-directions with 5 sample repetitions. To investigate the SW dynamics, we applied rf field excitation using a strip antenna of width 20nm at the centre of the sample.
The excitation was in the form of a temporally varying sinc pulse of the mathematical form $B(t)= B_{o} \text{sinc} (2\pi f_{max}(t-t_{o}))$, with amplitude $B_{o}=10$ mT, pulse offset $t_{o}=5$ ns, maximum cutoff frequency $f_{max}=30$ GHz and total evolution time t = 15 ns. Subsequently, the dynamic magnetization states $(m(r, t))$ were recorded for every 10 ps, which were further used to calculate the magnonic band structure as well as the spatial distribution of SW amplitude. For the calculation of the frequency vs. wave vector dependence, a 2D Fast Fourier Transform (FFT) was performed over the magnetization states after eliminating the final static contribution. To estimate the SW mode profiles for different frequencies, we first computed the FFT of the spatial average of the magnetization deviation $(m_{y}(t) - m_{y}(0))$ which determines the excitation spectrum. SW spatial mode profile for each resonant frequency was then determined by taking the corresponding FFT amplitude in each cell.
\begin{center}
    \begin{figure*}[t]
    \centering
    \includegraphics[width = 0.75\linewidth]{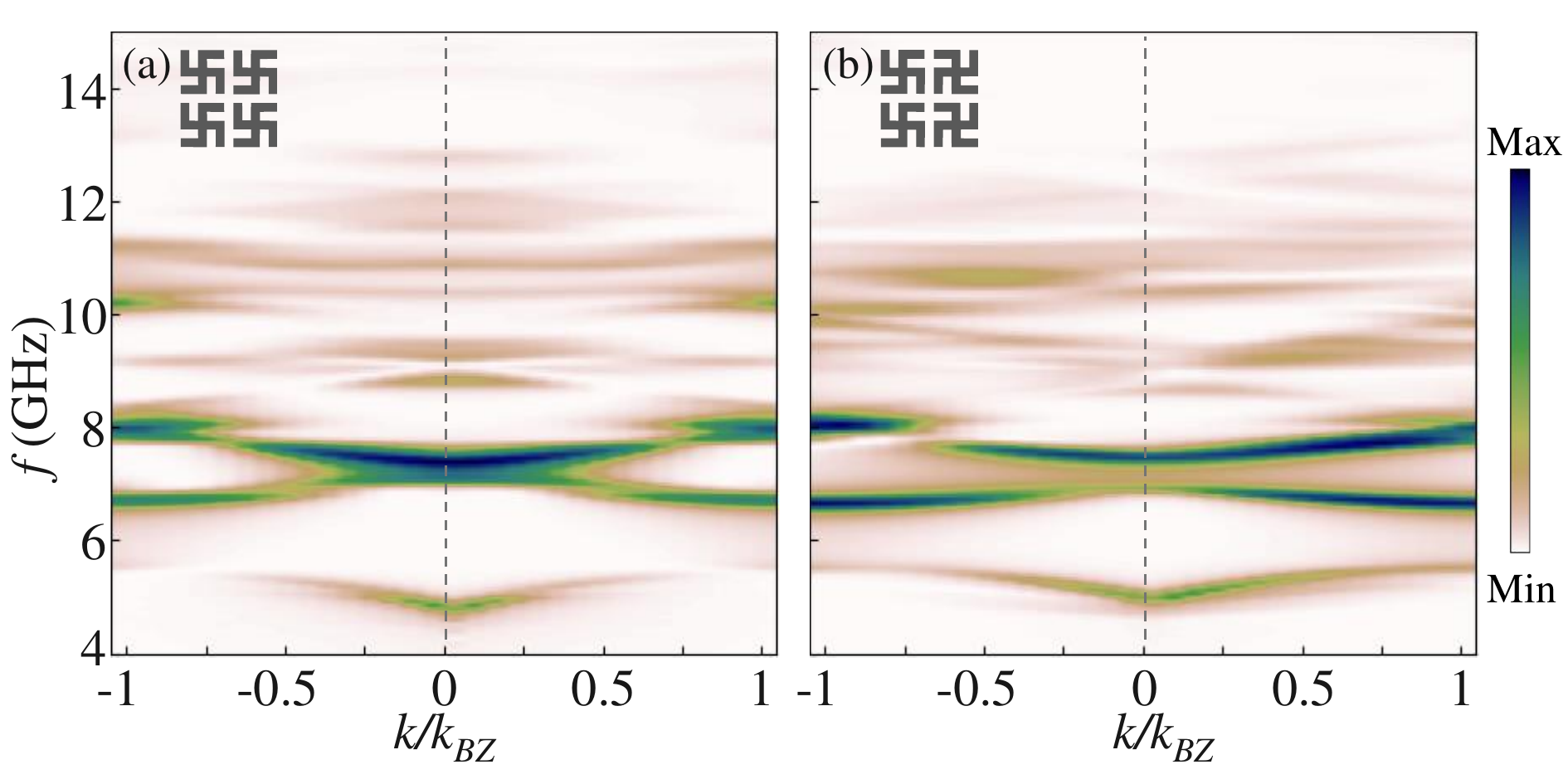}
    \caption{Simulated magnonic band structure up to $1^{\text{st}}$ Brillouin zone boundary for (a) CMC and (b) CMCAH. For CMC, the dispersion is symmetric for +\textit{k} and -\textit{k}, while for CMCAH, there is a non-reciprocity. Insets provide schematic of the respective arrangements of the enantiomers.}
    \label{FIG. 2.}
    \end{figure*}   
\end{center}
\maketitle
\section{Results and Discussions}
\subsection{ Magnetic configuration}
The final static magnetic configurations of both CMC and CMCAH are shown in Fig. \ref{FIG. 1.}b (left and right panels respectively). For obtaining the magnetic configurations we initially considered random spin orientations in the sample, followed by application of external bias magnetic field $B_{ext}=80$ mT along x-direction, after which the system was left to relax until it reached the minimum energy state. The magnetic dipolar shape anisotropy favours the in-plane alignment of magnetic moment along the length of each arm. As a result, the spins tried to align themselves in an orientation consistent with the handedness of the system, i.e., favouring horizontal orientation for horizontal arms $(m_{1}= \rightarrow) $ and either up or down $(m_{2}=\uparrow \text{or} \downarrow)$ for vertical arms, exhibiting a chiral magnetic spin ordering. This led to a deviation of the spins from the applied magnetic field direction in the vertical arms, as shown by the y component of magnetization $(M_{y})$ encoded in the colormap in Fig. \ref{FIG. 1.}b. Depending on the sample’s geometry, the modification in the local magnetization states changes the dipolar coupling between the two consecutive elements which eventually drives the SW dispersion properties as we will see later in this article.

\subsection{Non-reciprocal spin wave dispersion}
\begin{center}
    \begin{figure*}[t]
    \centering
    \includegraphics[width = \linewidth]{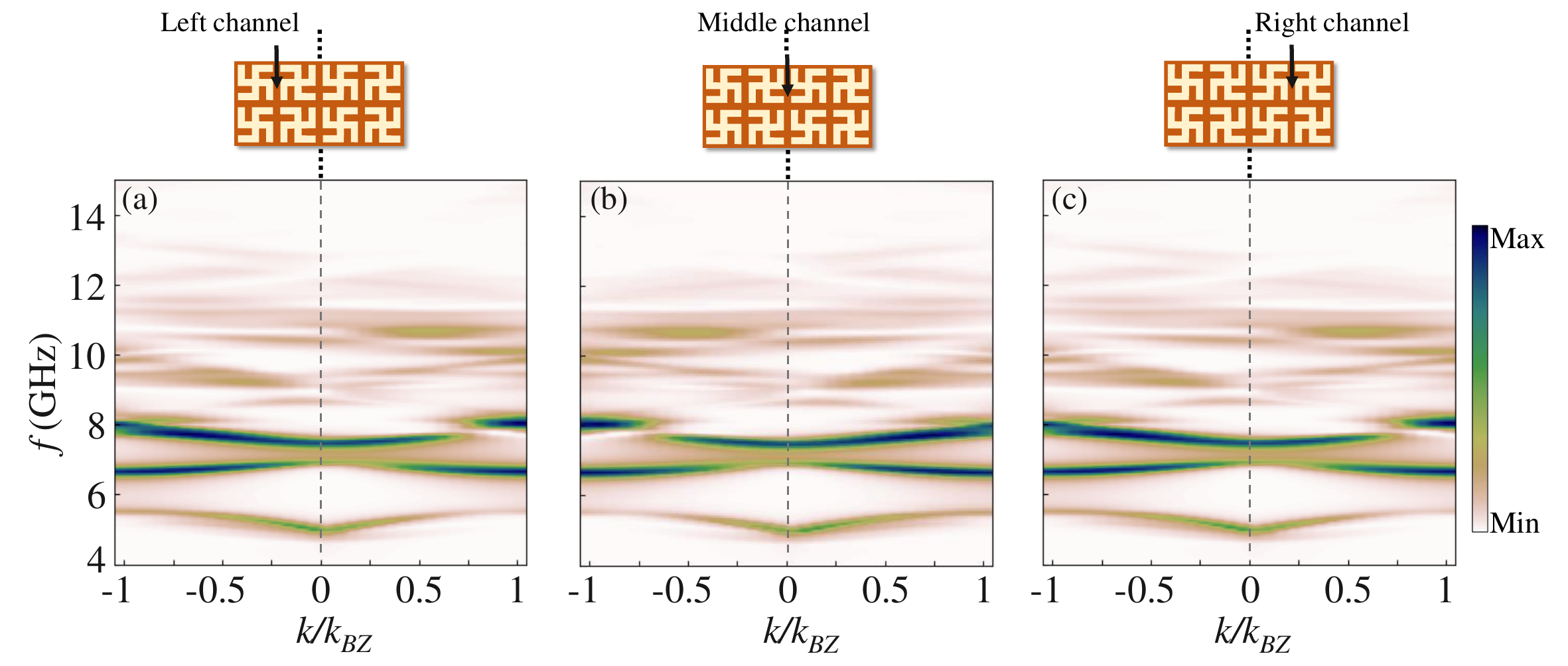}
    \caption{Channel selective SW nonreciprocity in CMCAH shown for (a) left channel, (b) middle channel and (c) right channel. The top panel provides visual representation of the left, middle and right channels respectively within the CMCAH.}
    \label{FIG. 3.}
    \end{figure*}   
\end{center}

Figure \ref{FIG. 2.} shows the SW frequency vs wave vector dispersion results up to $1^{\text{st}}$ Brillouin zone (BZ) boundary for both the MCs. To obtain the dispersion, 2D FFT was performed over the middle channel between two Py elements along the y-direction (See Fig. \ref{FIG. 1.}a and upper panel of Fig. \ref{FIG. 3.}b). Here the geometry is the so-called Damon Eshbach geometry \cite{r45_book_Anil_Prabhakar}, where the directions of $B_{ext}$ (x-axis) and the SW wave vector (y-axis) are perpendicular to each other, and the SWs are emitted on both sides from the antenna region (say positive wave vector +\textit{k} and negative wave vector -\textit{k} for the waves propagating in positive and negative y-directions respectively). \par The magnonic band structures of the CMC and CMCAH (Fig. \ref{FIG. 2.}a and \ref{FIG. 2.}b respectively) consist of a number of allowed and forbidden SW bands.The calculated group velocity of the mode near 7.5 GHz is approximately 300 m/s. However, a careful comparison of Fig. \ref{FIG. 2.}a and \ref{FIG. 2.}b points out some discernible changes in the dispersion characteristics as a result of symmetry breaking.
\vspace{-0.95cm} % Adjust the value as needed
First, an additional band gap appeared at the centre of the BZ as we moved from CMC to CMCAH. More importantly, the modes are symmetric with respect to +\textit{k} and -\textit{k} in CMC, whereas, in contrast, they exhibit a pronounced asymmetry (non-reciprocity) for CMCAH. \par These features can be described based on the dynamic magnetization characteristics evolving from the ground state spin configuration (Fig. \ref{FIG. 1.}b). In the CMC, the translational symmetry persists along both the x and y-directions. Consequently, the pattern of dipolar coupling for one element is identical to its neighbouring one, resulting in symmetric dispersion with respect to wave vector \textit{k}.

As we move to the other system, two interesting observations emerge. First, the equivalent arms of the left handed and right handed enantiomers are now antiparallel. Additionally, the symmetry breaking creates distinct environment for the counter propagating wave vectors along the y-axis. For example, if the +\textit{k} spin wave sees the left handed and right handed elements facing each other, for -\textit{k}, they will face away from each other. The associated dynamic dipolar interaction energy $( \textit{E}_d = -(\mu_{o}/2)m.h_{dip})$ is lower for the parallel orientation between stray fields and the dynamic magnetization. As a result, the SWs with +\textit{k} and -\textit{k} are no more equivalent, leading to the non-reciprocity in amplitude and frequency. In Appendix A, we have shown the magnetostatic field distribution for both the MCs. In case of CMC, the dipolar field profile across two consecutive columns looks identical with respect to +y and -y directions, which is not the case for CMCAH. This is clearer when the evolution of dynamic dipolar coupling is observed (see the videos in Supplementary Material [46] for dynamic dipolar field distribution after excitation of spin wave mode near 7.5 GHz and considering two consecutive columns of each MC). In case of CMC, the excitation propagates symmetrically on both sides of the antenna. However, a significant difference in the dynamic dipolar coupling strength can be observed for the structure with alternate handedness, meaning the dipolar coupling exhibits a pronounced asymmetry for +\textit{k} and -\textit{k}.

The chiral spin configuration along with the symmetry breaking also give rise to a band gap opening at around 7.2 GHz at the centre of the BZ, which can be attributed to the Bragg scattering of spin waves from the periodic structuring, which give rises to a periodicity in the dipolar field distribution. We verified this by simulating the magnonic band structure for varying dipolar coupling strength which is actually dependent on the inter-element separation \textit{d} (See inset of Fig. \ref{FIG. 1.}a). The results show that the band gap gets wider (closes) by decreasing (increasing) \textit{d} (See Appendix B), which confirms the origin of the bandgap lies in the dipolar coupling between the elements. This band gap is different from the indirect band gap observed earlier in MCs with periodic iDMI, where the band gap was formed because the dispersion curves of counter propagating waves had different slopes \cite{r41_experimental_flat_bands_1D_MC,r42_flatbands_indirect_bandgaps_induced_by_periodic_DMI}. \par If the above argument based on asymmetric dipolar coupling is true, then the SW non-reciprocity should be controlled by changing the relative orientation of the left and right handed elements, prompting an exploration of channel-selective SW propagation. In order to verify that, we selected three vertical YIG channels in the CMCAH as indicated in the top panel of Fig. \ref{FIG. 3.}. The left and the right channels have similar surroundings, where the left-handed and right-handed chiral elements face each other. In contrast, for the middle channel, these left-handed and right-handed elements face in opposite directions. The corresponding SW dispersion curves for +\textit{k} and -\textit{k} are depicted in Fig. \ref{FIG. 3.}. Strikingly, the directional asymmetry and non-reciprocal behaviour in the magnonic band structure completely reversed for alternate SW channels. This validates that the dipolar coupling between left and right handed elements have different effects on the SWs propagating in +y and -y directions which can be effectively tuned by changing their relative orientations. On contrary, the dispersion behaviour of CMC remained consistent across all channels. This flexibility of on-demand tuning of SW propagation by external patterning stands out as a remarkable effect, which is attributed to the structural chirality of the crystal.

\subsection{Channel selective propagation and tunability}
\begin{center}
    \begin{figure*}[t]
    \centering
    \includegraphics[width = \linewidth]{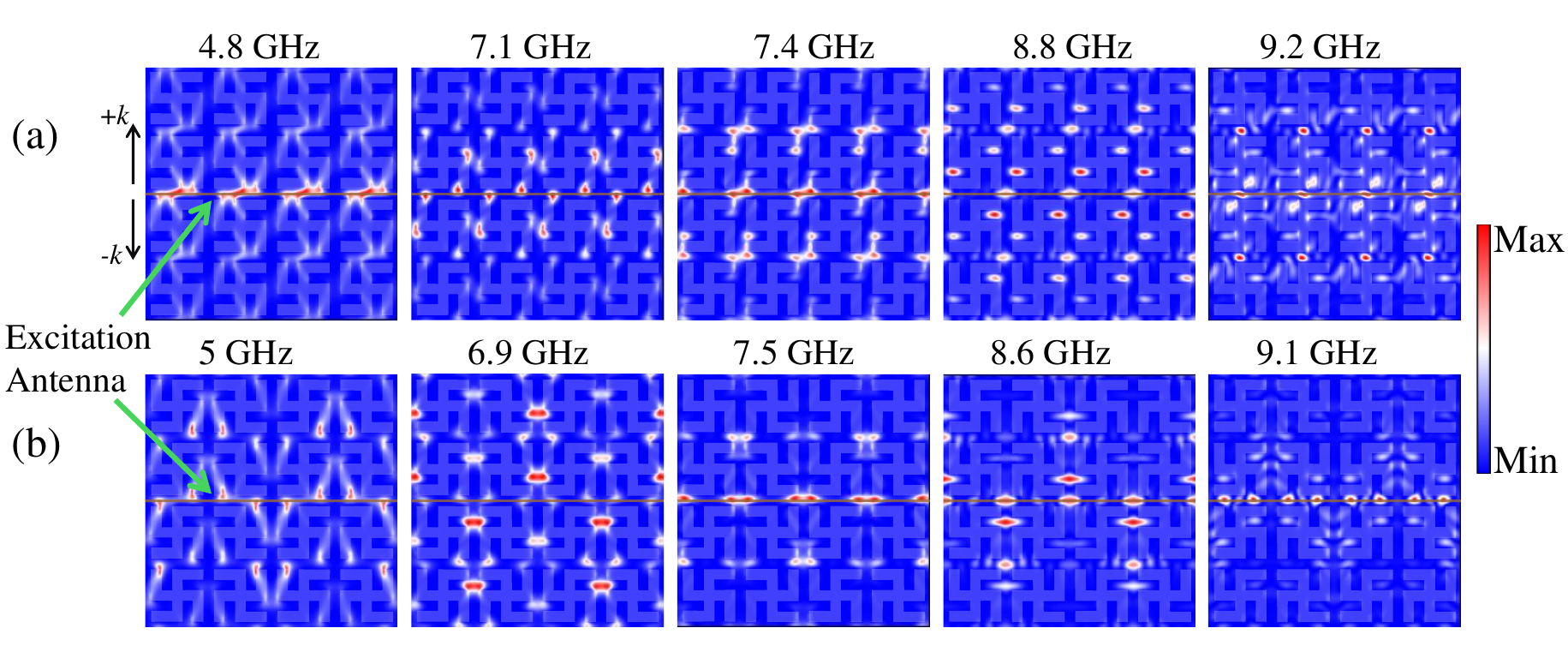}
    \caption{Spatial SW amplitude profiles for different frequencies in (a) CMC and (b) CMCAH. The excitation antenna with planar dimensions 1.2\textmu m$\times\text{20nm}$ is shown as a horizontal line in the middle. The locations of the chiral elements are indicated in grey shades within the mode profiles as a guide to eye.}
    \label{FIG. 4.}
    \end{figure*}   
\end{center}

In order to get closer insight into the directional properties of SWs and its tunability, we simulated the spatial amplitude distribution of the resonant SW modes, which are determined by taking FFT of the magnetization of the whole system. Fig. \ref{FIG. 4.} elucidates the spatial amplitude profiles for some SW modes in CMC and CMCAH (Fig. \ref{FIG. 4.}a and \ref{FIG. 4.}b respectively). For CMC and CMCAH these resonant frequencies slightly differ from each other due to differences in their structural arrangement and the corresponding distribution of demagnetized areas. For the profiles of the rest of the modes please refer to the Appendix C. In Fig. \ref{FIG. 4.}, the excitation antenna is shown by the straight line in the middle of the sample and snapshots of
\vspace{-0.5cm}
$1.2\mu\text{m}\times1.2\mu\text{m}$ sample dimensions are snipped for demonstration. This basically gives a visual map of the propagating SW modes for +\textit{k} and -\textit{k}. The periodic patterning leads to interference and scattering of the spin waves, resulting in quantized SW modes, which are extended along the y-axis perpendicular to the excitation antenna. This causes a reduction in the group velocity as evident from Fig. \ref{FIG. 2.} and Fig. \ref{FIG. 3.}. 

Despite the quantized nature, the mode characteristics are significantly different in CMC and CMCAH. As for the CMC, the modes persist in different regions in the crystal, whereas in CMCAH, the modes are primarily confined close to the vertical YIG channels between the left handed and right handed elements. Another substantial difference can be noticed in the relative SW intensity for +\textit{k} and -\textit{k}, as denoted by the colormap. In case of CMCAH, for a particular channel, the SW intensity is higher on one side of the antenna as compared to the other side, implying a non-reciprocal trait. For each SW mode, this non-reciprocity dramatically flips to the other wave vector direction for alternate channels. This is clearly not the case in CMC, where the SW mode profiles are perfectly symmetric with respect to +\textit{k} and -\textit{k}. On the other hand, one can also observe that the SW asymmetry profile in the CMCAH is strongly dependent on the frequency, which is attributed to the respective coupling of dynamic magnetization with the dipolar stray field. Note that, unlike the DMI based systems reported earlier \cite{r25_i-DMI_Pt/CoFeB_films,r46_iDMI_perpendiculary_magnetized_ultrathin_films,r47_asymmetric_SW_dispersion_due_to_DMI,r48_SW_nonreciprocity_AFM_coupled_multilayers_dipolar_and_iDMI}, the SW mode profiles for CMCAH remain unaltered upon reversing the direction of the applied magnetic field (see Appendix D). This intriguing constancy underscores the time reversal invariance of the system as well as contributes towards maintaining signal integrity during transmission.
\begin{figure}[h]
\centering
\includegraphics[width=0.75\columnwidth]{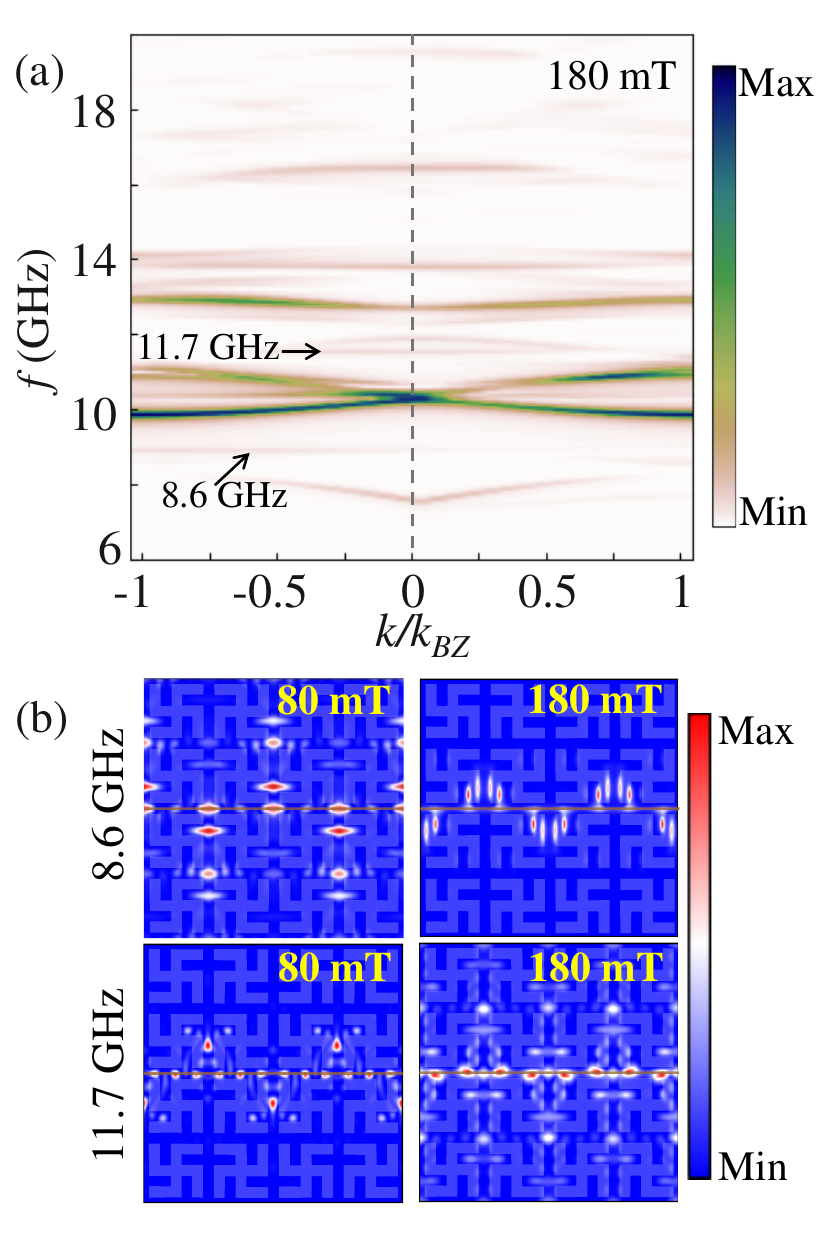}
\caption{(a) SW dispersion for CMCAH under an external bias field of 180 mT. (b) Demonstration of SW propagation tunability by varying magnetic field and SW frequency. The locations of the excitation antenna and the chiral elements follow Fig. \ref{FIG. 4.}.}
\label{FIG. 5.}
\end{figure}
\par The results presented in this work illustrate different facets of dynamic magnetic properties in CMCAH, signifying the non-reciprocal character of SWs and its tunability. We have also verified that the non-reciprocity still persists in backward volume geometry (See Appendix E), which facilitates the competence of this system for device application. Needless to say, a crucial requisite to attain the chiral magnonic properties and the non-reciprocity is the chiral spin texture which actually provides the necessary dipolar interaction. Hence a suppression of the spin texture will impact the SW non-reciprocity as well. We verified this by applying a higher magnetic field $B_{ext}=180\text{mT}$. The heightened $B_{ext}$ dominates over the chirality of CMCAH, mostly aligning the spins along the x-direction. The simulated band structure for this field is shown in Fig. 5a. At this magnetic field, the band gap completely closes and non-reciprocity in the dispersion diminishes considerably. These highlight the significance of the spin texture on the magnonic non-reciprocity which can be easily tailored by applied magnetic field. \par Magnonic crystals are familiar to the scientific community for more than a decade now for their potential applications in magnonic memory, sensors and spin-based signal-processing devices \cite{r13_SW_diode_and_circulator,r40_AC_tunable_MC,r49_MC_based_microwave_phase_shifters,r50_cavity_magnomechanics,r51_SW_computing_and_signal_processing,r52_magnonics_roadmap} by manipulating the SW dynamics. However, the capability to mould the SW direction just by in-plane structural engineering in a certain way was unexplored so far. The interplay between chiral symmetry breaking and magnon dynamics makes this system widely suitable for functional magnonic devices, such as resonant magnonic diode, a device that transmits spin waves incident from one direction and blocks the counter-propagating waves. The vertical SW channels between the enantiomers can be used for the same, where unidirectional SW transmission is permitted. The direction of wave communication can be changed by addressing neighbouring channels. As the mode propagation direction is further dependent on the frequency, propagation channel and resonant frequency selective SW communication can be realized. Another way to tune the transmission direction for SW signal processing is by adjusting the applied magnetic field. As mentioned earlier, an increase (decrease) of magnetic field diminishes (enhances) the chiral spin texture in these structures, impacting the non-reciprocity of the system.

\maketitle
To clarify, Fig. \ref{FIG. 5.}b shows the SW propagation for two frequencies, 8.6 GHz and 11.7 GHz for two $B_{ext}$ values, namely 80 mT and 180 mT. For these two frequencies two different functionalities are observed.
At 8.6 GHz, SW propagation in a certain channel is reversed by tailoring the magnetic field. On the other hand, at 11.7 GHz, one can tune between reciprocal and non-reciprocal SW character. In any case, the preferred transmission is protected by the sample geometry and is not affected by changing the direction of applied field, making it potentially competent for momentum locked magnonic waveguide.

\section{Conclusion}
In summary, using micromagnetic simulations, we present a novel approach to achieve broadband and tailorable SW non-reciprocity within a chiral magnonic crystal with alternate handedness without relying on DMI. The magnonic crystal consists of periodic array of artificially engineered chiral elements where the periodic arrangement of the enantiomers dictates the static and dynamic magnetic response of the system. Upon breaking the translation symmetry in one direction, the SWs propagating perpendicular to the symmetry breaking axis exhibit notable asymmetry in their amplitude and frequency for two wave vector directions, which can be effectively controlled by changing the relative orientation of the left handed and right handed enantiomers. Analysis of the magnonic band structure and the spatial amplitude profiles of different spin wave modes reveal broad tunability of the non-reciprocal characteristics via selection of propagation channel, resonant frequency and applied magnetic field. This asymmetry and non-reciprocity observed in SW modes are attributed to the chirality induced dipolar coupling through the arms of the elements, which can be externally tuned by the structural parameters of the crystal. The flexibility of dynamical tuning of the SW propagation is potentially useful for future magnon based information processing devices, such as magnonic diodes and isolators. Finally, the idea to induce magnonic chirality by engineering the shape of magnetic nanostructures not only opens a new direction in magnonics research, but also promises versatile spin wave manipulation, advancing future information processing and communication technologies.
\maketitle
\section{Acknowledgments}
D. P. gratefully acknowledges IIT Kanpur for providing financial support through Institute Fellowship. C. B. is thankful to IIT Kanpur for Research Initiation Grant.

\maketitle
\section{Appendix}

\subsection{Dipolar field distribution for the magnetic nanostructures}
The periodic magnetostatic field distributions for CMC and CMCAH are shown in Fig. A1. In case of CMC, due to the periodic arrangement of the chiral elements, the dipolar field distribution is identical across the neighbouring elements (Fig. A1 a). As for the CMCAH, the alternate arrangement of the enantiomers gives rise to asymmetric dipolar coupling with respect to +y and -y directions (Fig. A1 b).
\setcounter{figure}{0}
\renewcommand{\figurename}{FIG.}
\renewcommand{\thefigure}{A\arabic{figure}}
\begin{center}
    \begin{figure*}[t]
    \centering
    \includegraphics[width = 0.65\linewidth]{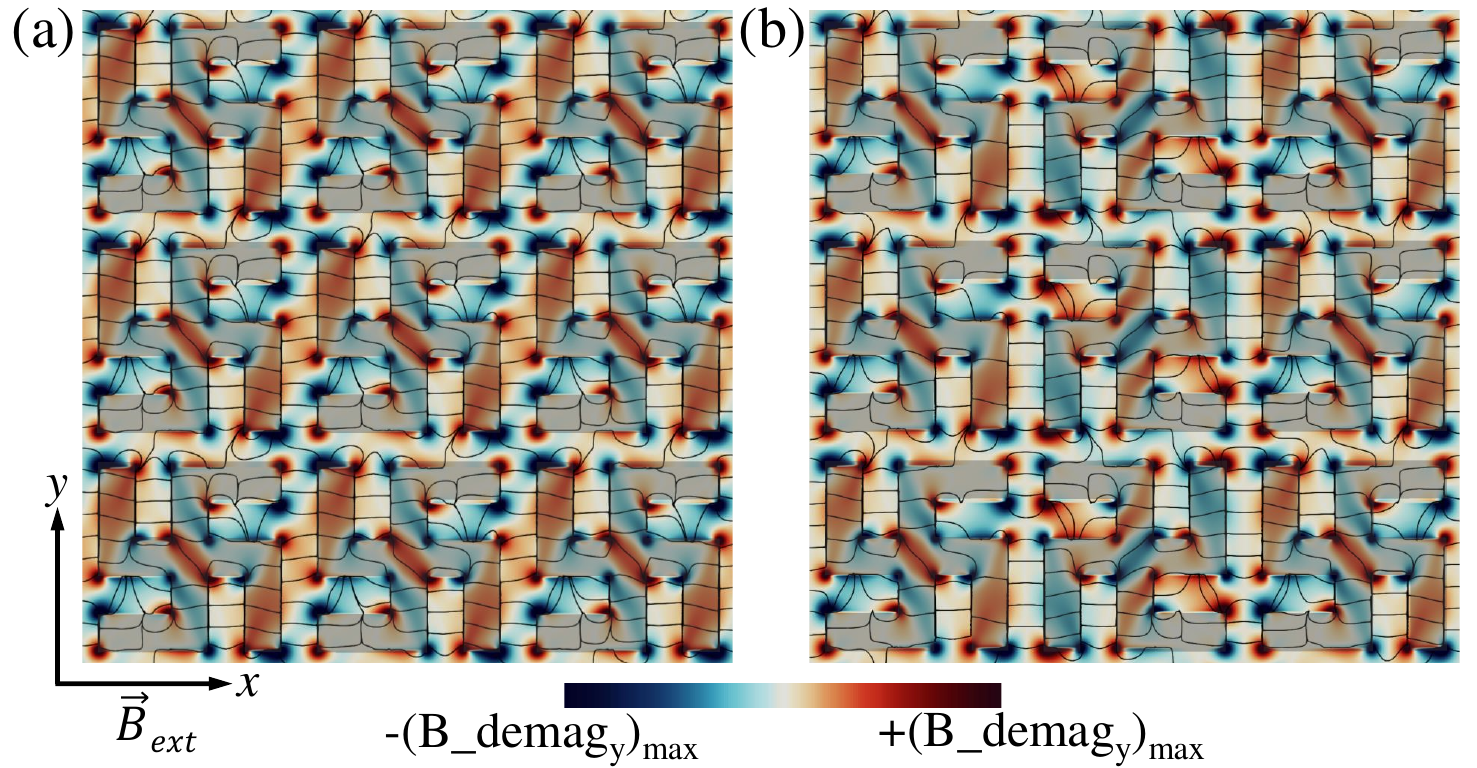}
    \caption{Magnetostatic field distribution for (a) CMC and (b) CMCAH. The locations of the chiral structures are indicated by the grey shades as a guide to eye.}
    \label{FIG. A1.}
    \end{figure*}   
\end{center}

\subsection{Spin wave dispersion by varying the separation between chiral elements in CMCAH}
\begin{figure}[h]
\centering
\includegraphics[width=\columnwidth]{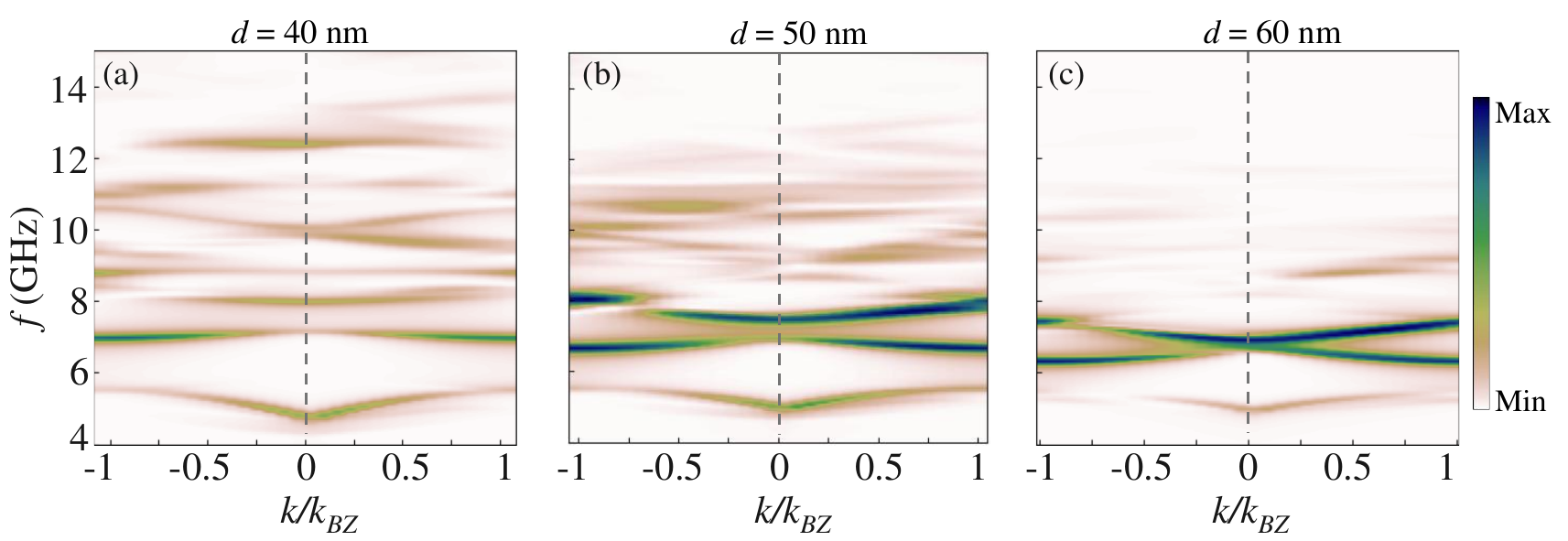}
\caption{Magnonic band structure of CMCAH for inter-element separation \textit{d} of (a) 40nm (b) 50nm (c) 60nm.}
\label{FIG. A2.}
\end{figure}
Magnonic band structure for different values of inter-element separation \textit{d} is shown in Figure A2. The results show that the width of the band gap occurring near 7.2 GHz are 1.06 GHz and 0.47 GHz for inter-element separation \textit{d} = 40nm and 50nm respectively. The band gap closes when \textit{d} is further increased to 60nm.

\subsection{Amplitude profiles of different spin wave modes}
For CMC the resonant modes at higher frequencies are shown in Fig. A3. The modes are quantized and a generic trend of symmetric SW propagation for +\textit{k} and -\textit{k} is observed, which is attributed to the same environment they are experiencing within the CMC.
\begin{figure}[h]
\centering
\includegraphics[width=0.75\columnwidth]{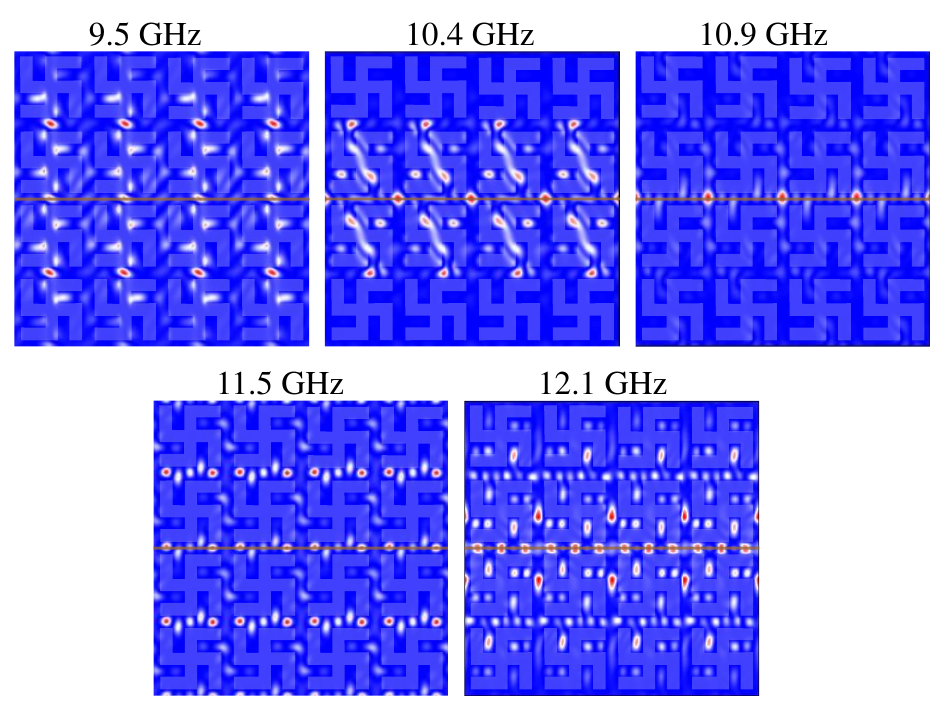}
\caption{Spin wave mode profiles at various frequencies for the CMC. The amplitude profiles were obtained by taking the FFT of dynamic y-component of magnetization $(m_{y})$. The excitation antenna is shown by the straight line in the middle. The locations of the chiral structures are indicated by the grey shades as a guide to eye.}
\label{FIG. A3.}
\end{figure}
\par Figure A4 shows the SW mode profiles in CMCAH for higher frequencies. For each YIG channel, the left and right-handed enantiomers face either toward or opposite to one another to create this broken symmetry. This configuration creates distinct environment for the counter propagating wave vectors along the y-axis, which leads to a non-reciprocity in the SW propagation in each channel. These results are consistent with our findings for the lower frequencies as shown in Fig. \ref{FIG. 4.}.

\begin{figure}[h]
\centering
\includegraphics[width=0.75\columnwidth]{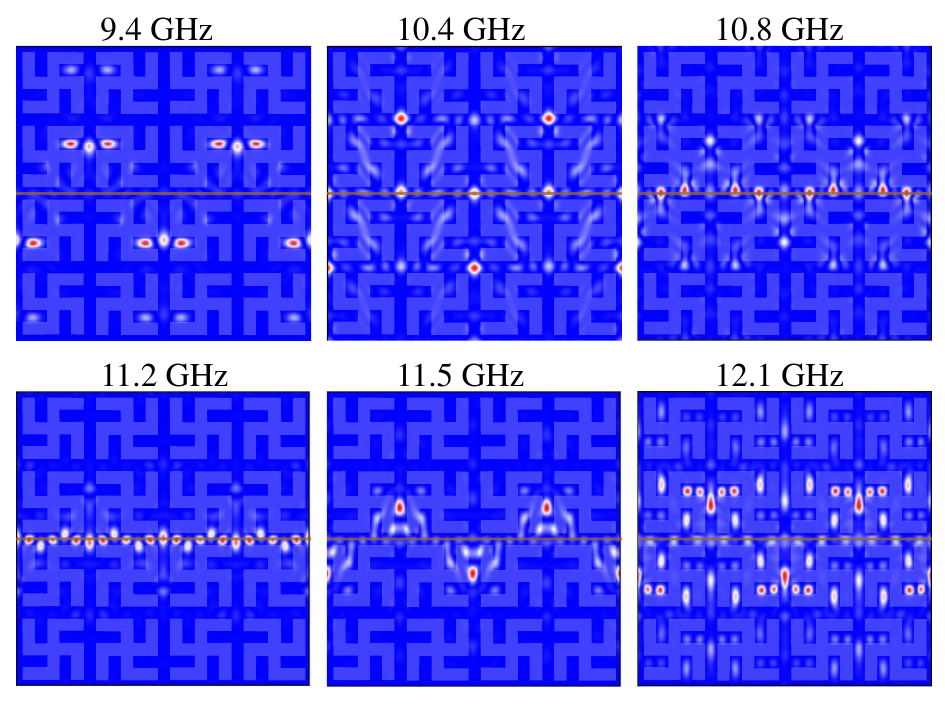}
\caption{Spin wave mode profiles for CMCAH at higher frequencies.}
\label{FIG. A4.}
\end{figure}

\subsection{Spin wave mode characteristics by reversing bias magnetic field direction}
We have investigated the magnonic band structure of chiral magnonic crystal with alternate handedness upon reversing the direction of applied magnetic field as shown in Fig. A5 a, b, c for the left, middle and right YIG channels respectively. Fig. A5 d presents respective spatial amplitude profiles for some SW modes. SW properties remains unchanged and are in complete accordance with Fig. \ref{FIG. 4.}. in the main text.
\begin{figure}[h]
\centering
\includegraphics[width=\columnwidth]{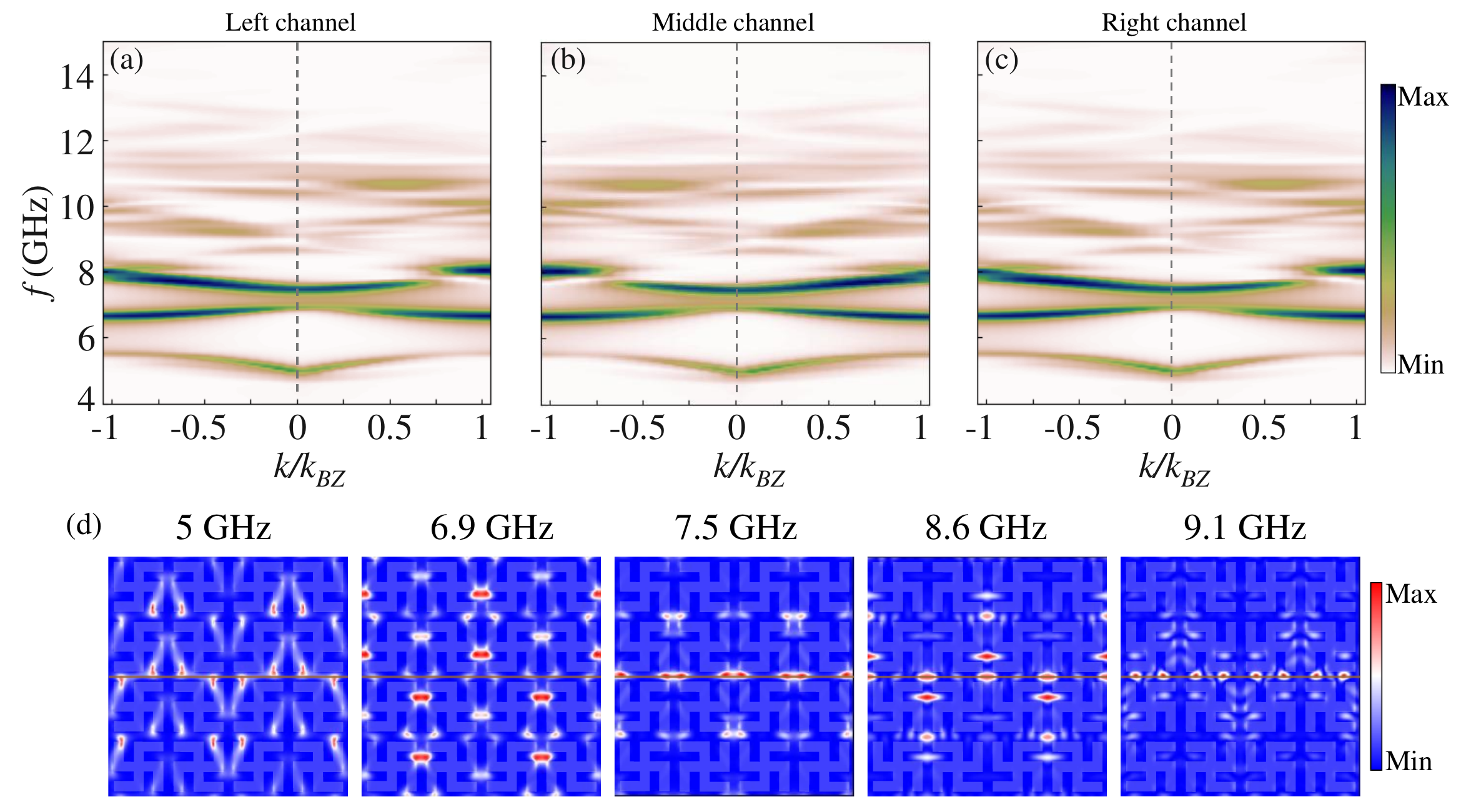}
\caption{Magnonic band structure of chiral magnonic crystal with alternate handedness for (a) Left channel (b) middle channel (c) right channel. (d) Spin wave mode profiles at resonant frequencies for CMCAH. The direction of bias field is along -ve x-axis following Damon Eshbach geometry.}
\label{FIG. A5.}
\end{figure}

\subsection{Simulation of SW Dispersion and spatial amplitude profiles for backward volume geometry}
We have investigated the magnonic band structure in the backward volume geometry (magnetic field parallel to wave vector direction) as shown in Fig. A6 a, b, c for the left, middle and right YIG channels respectively. Fig. A6 d presents the corresponding spatial amplitude profiles for some SW modes. Evidently, the SW non-reciprocity still persists, which facilitates the potential of this system as a SW diode device.
\begin{figure}[h]
\centering
\includegraphics[width =\columnwidth]{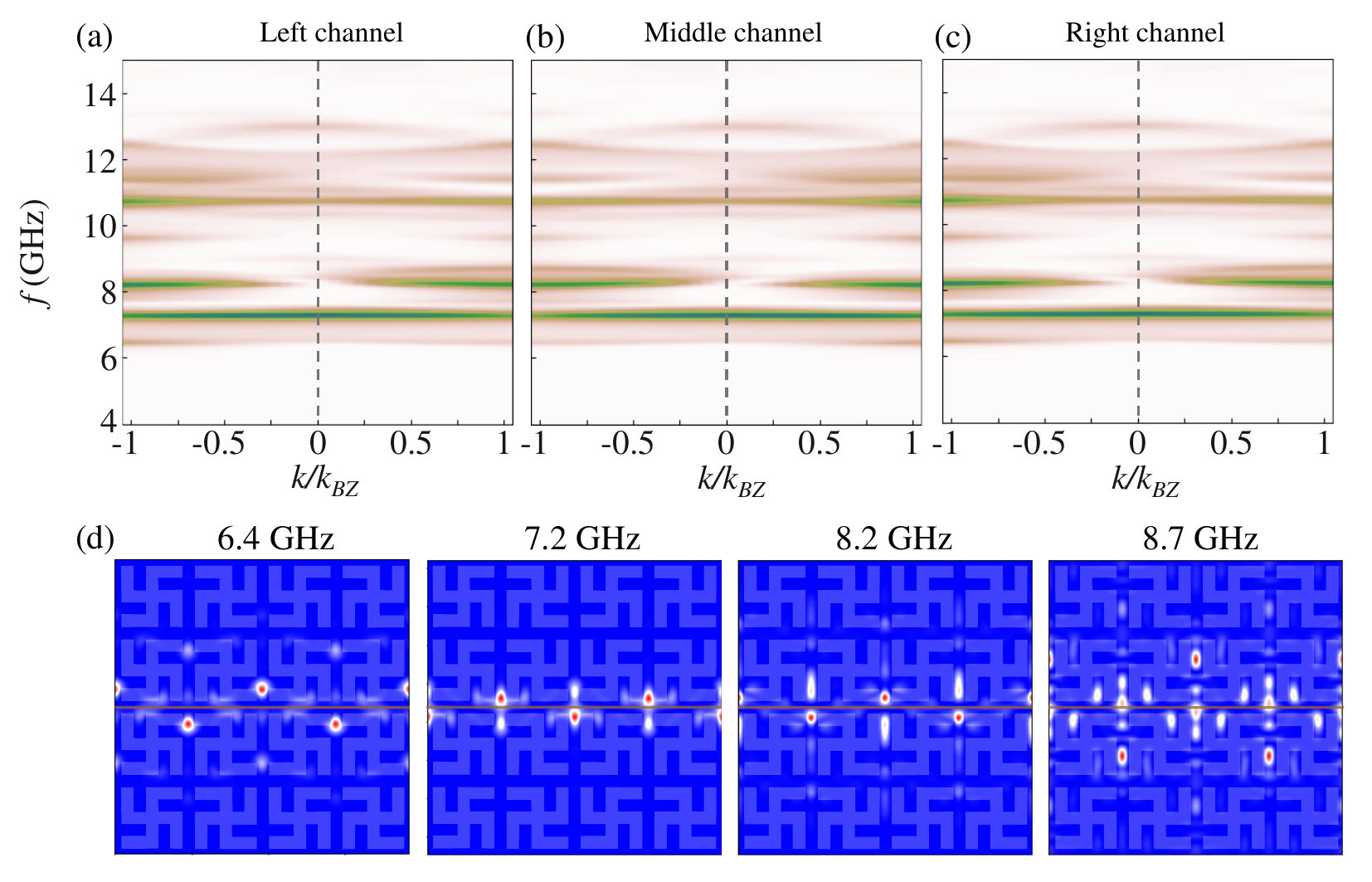}
\caption{Magnonic band structure of the CMCAH in the backward volume geometry for (a) left channel, (b) middle channel and (c) right channel. (d) Spatial SW amplitude profiles for some spin wave frequencies.}
\label{FIG. A6.}
\end{figure}
  
\newpage
%\bibliography{Ref}% Produces the bibliography via BibTeX.\\\\\\

\end{document}